\documentclass[
 aip,
 jap,%
 amsmath,amssymb,
reprint,%
final
]{revtex4-1}
\usepackage{graphicx}
\begin{document}

\title{Investigation of TiO$_x$ barriers for their use in hybrid Josephson and tunneling junctions based on pnictide thin films}

\author{S. D\"oring}
\email{sebastian.doering.1@uni-jena.de}
\author{M.~Monecke}
\author{S.~Schmidt}
\author{F.~Schmidl}
\author{V.~Tympel}
\affiliation{Friedrich-Schiller-Universit\"at Jena, Institute of Solid State Physics, Helmholtzweg 5, 07743 Jena, Germany}%
\author{J.~Engelmann}
\author{F.~Kurth}
\author{K.~Iida}
\affiliation{IFW Dresden, Institute for Metallic Materials, Helmholtzstrasse 20, 01171 Dresden, Germany}
\author{S.~Haindl}
\thanks{now at: Physikalisches Institut, Experimentalphysik II, Universit\"{a}t T\"{u}bingen, Auf der Morgenstelle 14, 72076 T\"{u}bingen, Germany}
\affiliation{IFW Dresden, Institute for Solid State Research, Helmholtzstrasse 20, 01171 Dresden, Germany}
\author{I.~M\"onch}
\affiliation{IFW Dresden, Institute for Integrative Nanosciences, Helmholtzstrasse 20, 01171 Dresden, Germany}
\author{B.~Holzapfel}
\thanks{now at: Institute for Technical Physics, Karlsruhe Institute of Technology, Hermann-von-Helmholtz-Platz 1, 76344 Eggenstein-Leopoldshafen, Germany};
\affiliation{IFW Dresden, Institute for Metallic Materials, Helmholtzstrasse 20, 01171 Dresden, Germany}%
\author{P.~Seidel}
\email{paul.seidel@uni-jena.de.}
 \affiliation{Friedrich-Schiller-Universit\"at Jena, Institute of Solid State Physics, Helmholtzweg 5, 07743 Jena, Germany}%
 
\begin{abstract}
We tested oxidized titanium layers as barriers for hybrid Josephson junctions with high $I_cR_n$-products and for the preparation of junctions for tunneling spectroscopy. For that we firstly prepared junctions with conventional superconductor electrodes lead and niobium, respectively. By tuning the barrier thickness we were able to change the junctions' behavior from a Josephson junction to tunnel-like behavior applicable for quasi-particle spectroscopy. Subsequently, we transferred the technology to junctions using Co-doped BaFe$_2$As$_2$ thin films prepared by pulsed laser deposition as base electrode and evaporated Pb as counter electrode. For barriers with a thickness of 1.5\,nm we observe clear Josephson effects with $I_cR_n$\,$\approx$\,90\,$\mu$V at 4.2\,K. These junctions behave SNS'-like and are dominated by Andreev reflection transport mechanism. For junctions with barrier thickness of 2.0\,nm and higher no Josephson but SIS'- or SINS'-like behavior with a tunnel-like conductance spectrum was observed. 
\end{abstract}



%74.25.F- transport properties of superconductors
%74.45.+c Proximity effects; Andreev reflection; SN and SNS junctions
%74.50.+r Tunneling phenomena; Josephson effects
%74.70.Xa Pnictides and chalcogenides 
%85.25.-j Superconducting devices

\maketitle
%\twocolumn

\section{Introduction}
Josephson junctions are a potent tool for the investigation of iron pnictide superconductors. In our investigations on iron pnictide superconductors we prepared planar Josephson junctions using a sputtered insulating layer as barrier. Up to now different works for Josephson junctions with single crystals, \cite{Zhang2009a,Zhang2009b,Zhang2012} for bi-crystal grain boundary junctions \cite{Lee2009,Katase2010a,Katase2011,Hiramatsu2012,Schmidt2013} and edge-type junctions \cite{Doering2012a, Doering2012c} as well as for planar junctions \cite{Schmidt2010} were published. The Josephson junctions presented therein used natural or engineered interfaces as well as normal metal layers as barrier, but up to now, no work with a deposited insulating barrier has been published. Additionally planar homo junctions with ferromagnetic barrier layers are in progress.\cite{Engelmann2013}
The behavior of a Josephson junction is most sensitive in the barrier properties e.g. its transparency. Our early experiments resulted in low resistance Josephson junctions with gold barriers showing an $I_cR_n$-product of about 18\,$\mu$V, \cite{Schmidt2010} which is much too low for possible applications like SQUIDs. Other works mentioned above and summarized in Ref.~\onlinecite{Seidel2011} showed $I_cR_n$-products of similar magnitude for different kinds of Josephson junctions with iron pnictides and various barriers. Only for junctions with a native barrier $I_cR_n$-products up to 300\,$\mu$V could be observed \cite{Zhang2009b}.\\
One of the most famous phase-sensitive tests for the expected s$_\pm$-symmetry of iron pnictides from Parker and Mazin \cite{Parker2009} requires two distinct Josephson junctions with different transparencies. Thus, the transport of each junction will be dominated by the opposite charge carrier type so they will form a $\pi$-SQUID. Hence, it is obvious to reach the idea by the combination of two junctions with a metal-like (high transparency) and insulating (low transparency) barrier, respectively.\\
Additionally, previous junctions with gold barriers showed non-ideal features like high excess currents, asymmetric $V$-$I$ characteristics and non-constant background in microwave dependence of $I_c$. \cite{Schmidt2010} By changing the barrier from normal conducting gold to insulating titanium oxide, an increase of the $I_cR_n$-products of our planar junctions is highly expected and a reduction of these non-ideal effects seems to be possible.\\
Another reason to introduce insulating barriers would be their use in quasi-particle spectroscopy on iron-pnictides. In low resistance junctions an additional series resistance from the electrodes can occur, when the measurement is done in the range of the transition temperature or with currents above $I_c$ of the film or electrode, respectively. \cite{Woods2004,Sheet2004,Baltz2009,Chen2010} This spreading resistance can dramatically disturb the observed conductance spectra or at least impedes their normalization necessary for modeling within the BTK-theory. \cite{Blonder1982} The effect of spreading resistance was observed in former junctions \cite{Schmidt2012} as well as in Point Contact Andreev Reflection (PCAR) studies on pnictides. \cite{Gonnelli2009b, Gonnelli2009c} Even if its possible to recalculate the real spectra with a semi-empiric model by characterization of the used electrodes, \cite{Doering2013b} this means more expenditure and results only in a rough estimation. By using insulating barriers, it should be possible to produce junctions with much higher resistances. Thus spreading resistances from the electrodes should be negligible in these junctions.\\
For conventional superconductors titanium oxide barriers have been used to prepare Josephson or SIN-junctions. \cite{Celaschi1985,Otto2007} Thus prior to the process on pnictide thin films, we used the barriers on junctions with two conventional superconductors lead and niobium, respectively, to test the barrier properties with well known materials.\\

\section{Junctions with conventional electrodes}
\label{Nb-junctions}

\begin{figure}[htb]
\centering
\includegraphics[width=0.95\columnwidth]{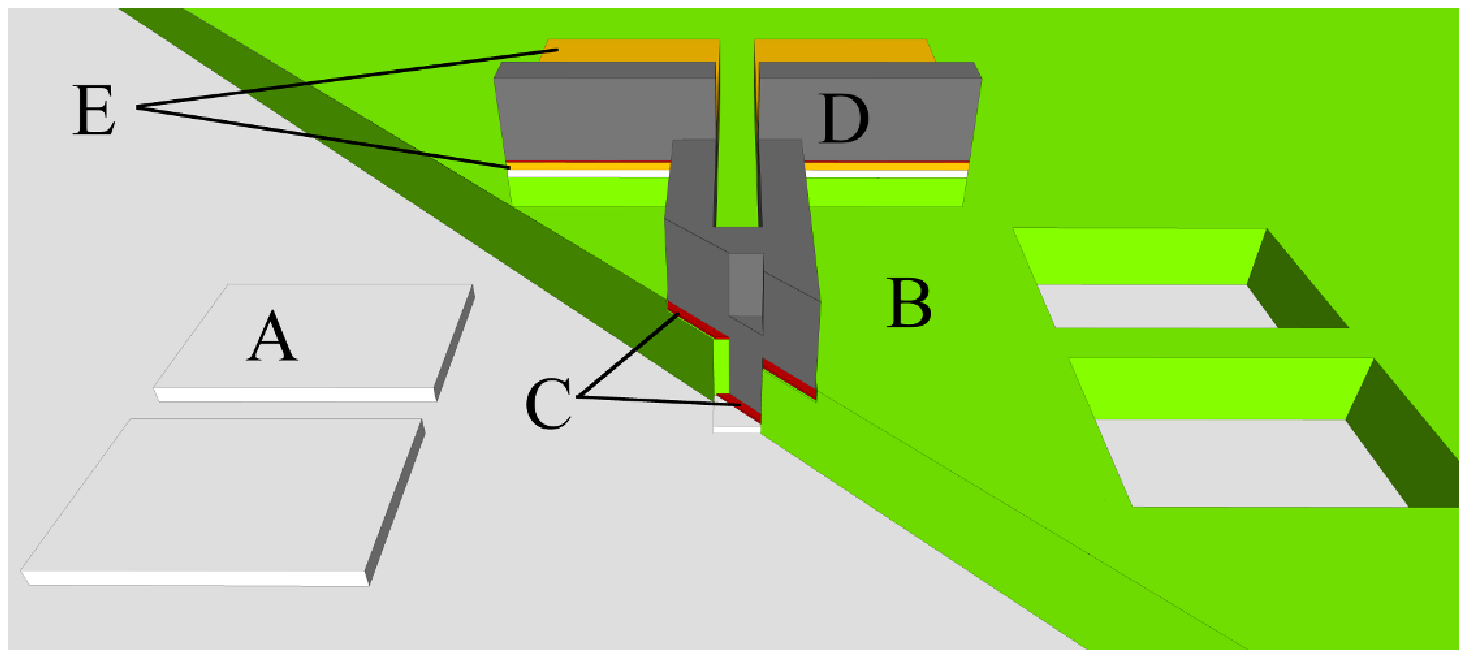}\\[6pt]
\includegraphics[width=0.95\columnwidth]{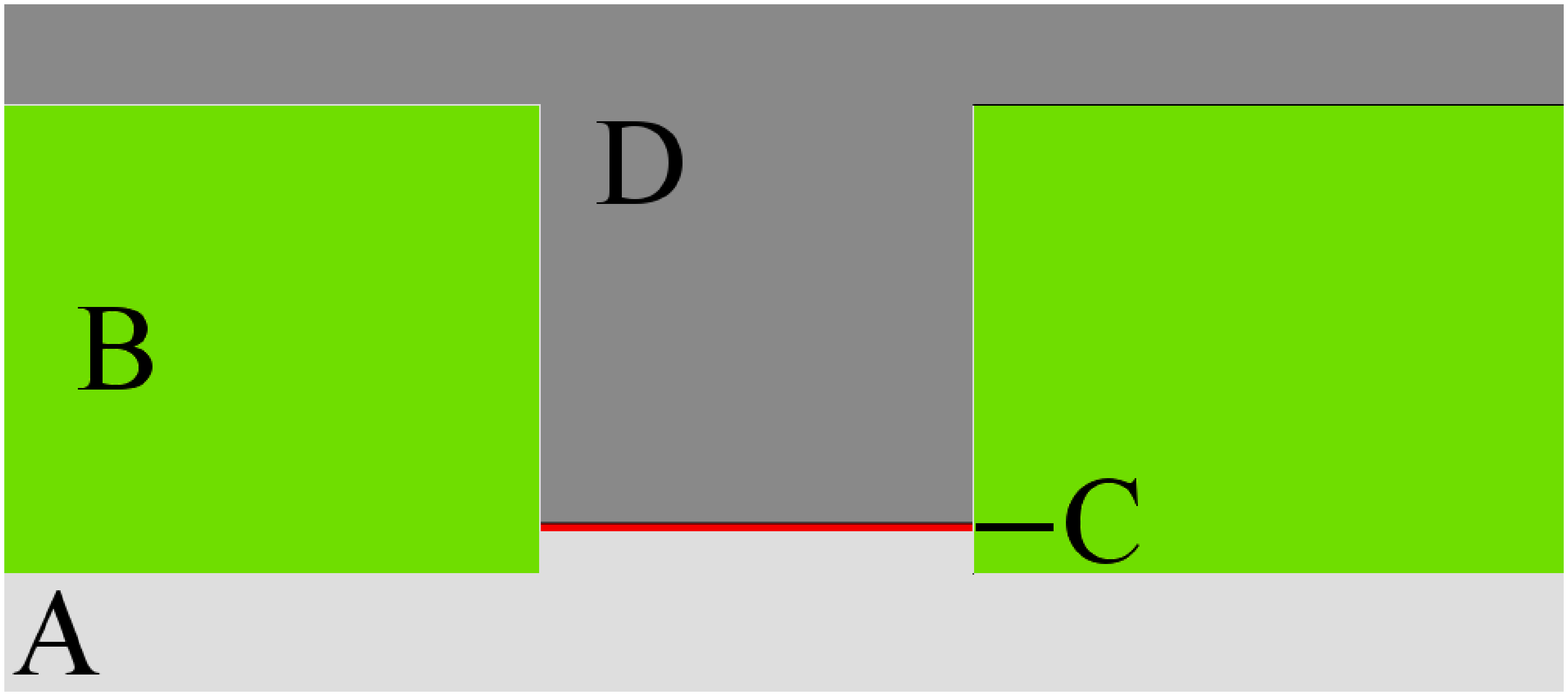}
\caption{\label{fig:Kontakt_Nb}Top: 3D sketch of a junction consisting of A(light grey) Nb base electrode, B(green) SiO$_2$ insulation, C(red) TiO$_x$ barrier, D(dark grey) Pb counter electrode and E(white\textbackslash yellow) Ti\textbackslash Au bonding pads. For a better view, all parts upside of the base electrode are cutted diagonal over the junction area.  Bottom: Cross-section of a junction.}
\end{figure}

\begin{figure}[htb]
\centering
\includegraphics[width=.95\columnwidth]{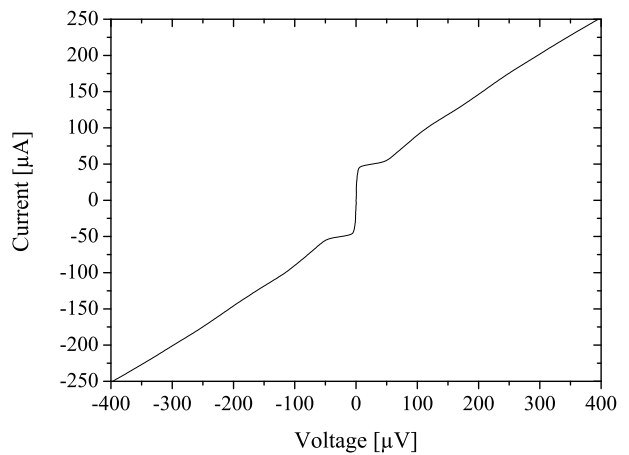}
\caption{\label{fig:Josephson_Nb}$I$-$V$-characteristic of a Nb\textbackslash TiO$_x$\textbackslash Pb junction with $d$(TiO$_x$)\,=\,2.5\,nm and an area of 10\,$\mu$m$\times$\,10\,$\mu$m at $T$\,=\,4.2\,K.}
\end{figure}

For the preparation of these Nb\textbackslash TiO$_x$\textbackslash Pb junctions we use a SiO$_2$ substrate with an e-beam evaporated niobium layer of 500\,nm thickness. To prepare the junction area and the bonding pads of the Nb electrode we use a photo-lithography mask and subsequently ion beam etching (IBE) to remove material from the Nb surface of 30\,nm depth. After this we deposit SiO$_2$ by reactive sputtering with a thickness of 250\,nm. The next step is the preparation of our counter electrode bonding pads with another photo-lithographic mask. For the pads we use a bi-layer system of titanium and gold with a thickness of 40\,nm and 90\,nm, respectively. The titanium is very adhesive on the sputtered SiO$_2$ and gold acts very well as electrical and mechanical contact for future gold wire bonding. Both materials are sputtered \textit{in-situ} at a pressure of 2\,Pa in argon atmosphere and with a power of 50\,W. After this we apply a photo-lithographic mask on the sample to deposit the barrier layer of oxidized Ti and the counter electrode layers of Pb and In. For the preparation of the barrier layer we clean up the Nb surface in the junction area from any oxides by removing 15\,nm of the surface via IBE. Thereafter the titanium is deposited \textit{in-situ} on the clean surface in Argon atmosphere at a pressure of 4\,Pa and a sputtering power of 30\,W at a deposition rate of 2\,nm/min. Thus we have good control of the barrier thickness. We heat the sample in air up to 80\,\symbol{23}C for 20\,min to get a fast and controlled oxidation of the titanium. For the deposition of the counter electrode we use vacuum evaporation of lead with a layer thickness of 350\,nm and followed by an indium protective layer with a thickness of 50\,nm. A 3D-sketch and the corresponding cross-section of this junction is shown in fig.~\ref{fig:Kontakt_Nb}.
\\
An $I$-$V$ measurement for this junction type with a barrier thickness $d$(TiO$_x$)\,=\,2.5\,nm is shown in fig.~\ref{fig:Josephson_Nb}. One can see RSJ-like behavior with $I_c\approx$\,50\,$\mu$A and $R_n\approx$\,2.0\,$\Omega$.
\begin{figure}[htb]
\centering
\includegraphics[width=.95\columnwidth]{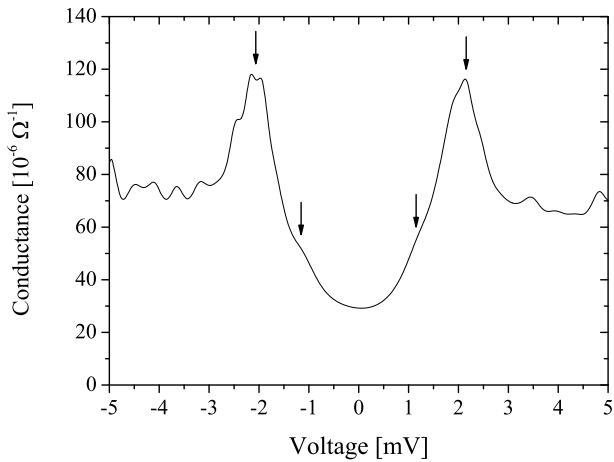}
\caption{\label{fig:spectrum_Nb}Conductance spectrum of a Nb\textbackslash TiO$_x$\textbackslash Pb junction with $d$(TiO$_x$)\,=\,3.0\,nm and an area of 15\,$\mu$m$\times$\,15\,$\mu$m at $T$\,=\,4.2\,K.}
\end{figure}
By increasing the barrier thickness it is possible to increase the resistance of the junction noticeably. In fig.~\ref{fig:spectrum_Nb} one can see a tunnel-like spectrum at $T\,=\,4.2\,$K with a normal state conductivity of 75\,$\cdot$10$^{-6}$\,$\Omega^{-1}$ or a resistance of 13.3\,k$\Omega$, respectively. The spectrum, which was measured for a junction with $d$(TiO$_x$)\,=\,3.0\,nm, shows peaks at $\vert eV\vert\,=\,\Delta_1\,+\,\Delta_2\,=\,2.2$\,meV and inner shoulders at $\vert eV\vert\,=\,\Delta_1\,\approx\,\Delta_2\,=\,1.1$\,meV, which are both marked by arrows in fig.~\ref{fig:spectrum_Nb}. These values are a little bit lower than expected for ideal materials for this temperature. For all measured junctions and the subsequently analyzed spectra the fitted gap values were 1.04\,meV\,$\leq\Delta_{Pb}(0)\leq$\,1.16\,meV and 1.23\,meV\,$\leq\Delta_{Nb}(0)\leq$\,1.36\,meV, respectively, while the ideal values are 1.35\,meV for lead and 1.45\,meV for niobium. Thus a small degradation of the superconducting electrodes at the barrier interfaces can be assumed.\cite{Seidel1980}\\

\section{Junctions with pnictide electrode}
\begin{figure}[htbp]
\centering
\includegraphics[width=.95\columnwidth]{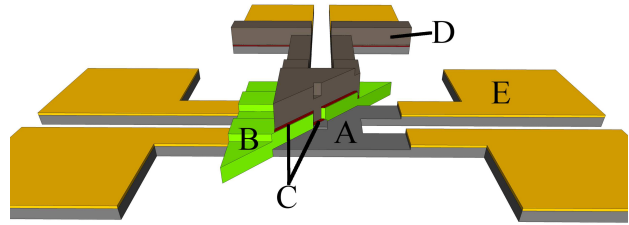}
\caption{\label{fig:Kontakt_122}3D-sketch of a junction with A(yellow) Ba-122 base electrode, B(green) SiO$_2$ insulating framework, C(red) TiO$_x$ barrier, D(dark grey) Pb counter electrode and E(yellow) gold cover layer.}
\end{figure}

The preparation of the Ba-122\textbackslash TiO$_x$\textbackslash Pb junctions follows in principal our well established process for former junctions with gold barrier, which can be found in detail in Ref.~\onlinecite{Doering2012b}. A first difference is the type of substrate we use for our pnictide thin film preparation. Instead of (La,Sr)(Al,Ta)O$_3$, we use CaF$_2$, which results in higher $T_c$ of the Co-doped BaFe$_2$As$_2$ (Ba-122) film. \cite{Kurth2013} Then we follow the processing steps from Ref.~\onlinecite{Doering2012b}, covering the whole thin film with a 10\,nm gold layer, pattering of base electrode via IBE, and forming the junction area by IBE and sputtering SiO$_2$ frameworks. Now we use a separate photo-lithographic mask, which covers the whole sample except an area including each junction window plus some tolerance area around. Via IBE the covering gold thickness is reduced to nominally 1\,nm. The reason for that is to reduce the influence of gold maximally, but not etch into the superconducting layer to avoid possible destruction of the superconductivity on the surface. Subsequently, we use a manipulator to transfer the sample from the IBE chamber into the sputtering chamber without breaking the vacuum. We sputter Ti layers with thicknesses between 1.0 and 3.0\,nm. After the sputtering process, the sample is tempered and oxidized in atmosphere like described above. Finally, we continue with our usual step and evaporate a Pb film for the counter electrode followed by an In layer as protection. A sketch of this junction is shown in fig.~\ref{fig:Kontakt_122}.
\\
In this work we will present results and show differences of samples with 1.5\,nm and 2.0\,nm of sputtered and oxidized titanium. Between these values there seems to be the transition from metal like to insulating behavior of the barrier. For barriers thinner than 1.0\,nm we cannot exclude short cuts due to incomplete covering of the superconducting surface, while thicknesses above 3.5\,nm result in resistances in the order of M$\Omega$.\\
\subsection{Josephson junctions}
\begin{figure}[htbp]
\centering
\includegraphics[width=.95\columnwidth]{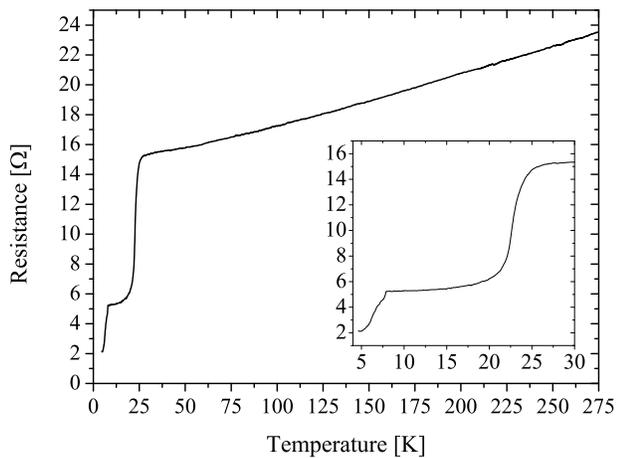}
\caption{\label{fig:R-T_A1} Resistance vs. temperature of a Ba-122\textbackslash TiO$_x$\textbackslash Pb junction junction with $100\,\mu$m~$\times~100\,\mu$m area and $d$(TiO$_x$)\,=\,1.5\,nm. The inset shows the same curve zoomed to the range between 4\,K and 30\,K, where both electrodes become superconducting.}
\end{figure}

\begin{figure}[htbp]
\centering
\includegraphics[width=.95\columnwidth]{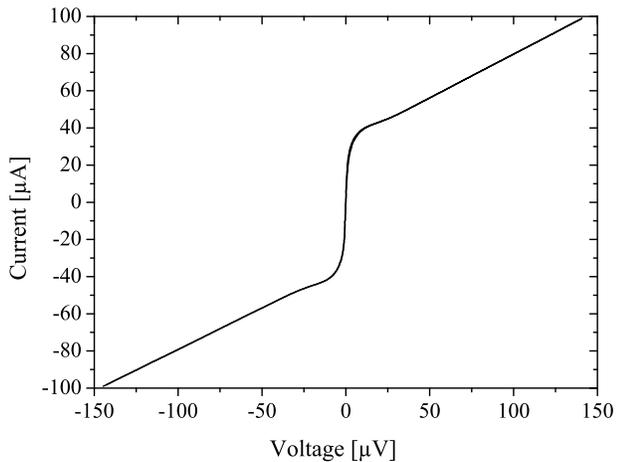}
\caption{\label{fig:V-I} Current vs. voltage characteristics and differential resistance vs. current of the same junction like in fig.~\ref{fig:R-T_A1} at $T$\,=\,4.2\,K.}
\end{figure}

\begin{figure}[htbp]
\centering
\includegraphics[width=.95\columnwidth]{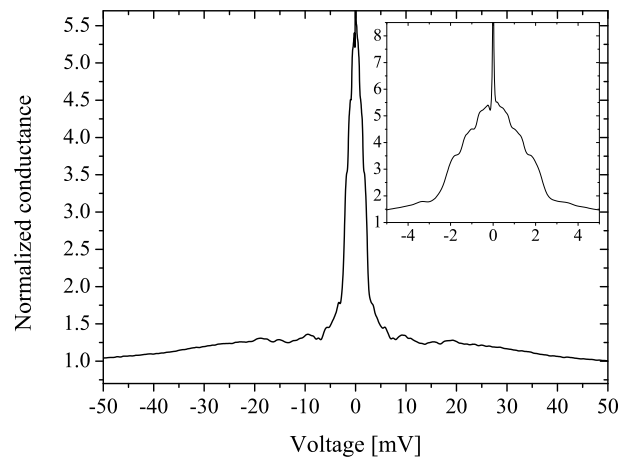}
\caption{\label{fig:G-U} Normalized differential conductance vs. voltage of the junction used in fig.~\ref{fig:V-I}. The inset shows the same spectrum in the range of low voltages.}
\end{figure}

The first junction we present has a barrier of 1.5\,nm thickness. As it can be seen in fig.~\ref{fig:R-T_A1}, the temperature dependence of the resistance shows linear behavior with small saturation above the superconducting transition of Ba-122 at $T_{on}\approx$\,26\,K. This is typical for a metal-like barrier, thus we can assume a SNS'-junction has been formed by such thin barrier. Nevertheless, this junction shows Josephson effect which differs from our former ones with a gold barrier. \cite{Schmidt2010} Firstly, the $I$-$V$-characteristic (see fig.~\ref{fig:V-I}) is not RSJ-like \cite{McCumber1968,Lee1971} but more similar to those Katase et al. observed for 16\,$^{\circ}$ grain boundary junctions. \cite{Katase2011} They show nearly linear behavior for $\vert I\vert>I_c$. Opposite to the gold barrier junctions the $I$-$V$ characteristic is nearly symmetric, in both branches equal values of $I_{c}\approx$\,40\,$\mu$A and $R_{n}\approx$\,2.2\,$\Omega$ can be estimated. This results in an $I_cR_n$ of nearly 90\,$\mu$V which is higher than for the gold barrier junctions \cite{Schmidt2010} by a factor of 5 and at least 1.5 higher than for grain boundary junctions \cite{Hiramatsu2012}, respectively. A detailed characterization of the prepared Josephson junctions and possible optimization of the barrier properties will be done in future work.
\\
By changing the measurement setup from resistance versus current to conductance versus voltage one can obtain information about the order parameter of the used superconductors. In fig.~\ref{fig:G-U} the conductance spectrum for the same junction as used for fig.~\ref{fig:V-I} is shown. One can see a central peak for $\vert V\vert\leq2.5$\,mV, with additional shoulders (see inset of fig.~\ref{fig:G-U}). Assuming a non-nodal order parameter this is a typical sign for transport dominated by Andreev-reflection and thus for a metallic barrier. Furthermore, some symmetric structures occur outside this central peak up to 20\,mV in the conductance, which are presumable due to phonon interactions. Such features were also observed in the former PCAR studies \cite{Tortello2010,Tortello2012} and are a footprint of the electron-boson spectral function $\alpha^2F(\Omega)$. For the case of pnictide superconductors, they can be interpreted within a three-band Eliashberg model. \cite{Ummarino2011} For even higher voltages there is a parabolic decreasing background, which can be explained by Joule heating of the junction. \cite{Baltz2009}\\

\subsection{Tunneling junctions}
\begin{figure}[htbp]
\centering
\includegraphics[width=.95\columnwidth]{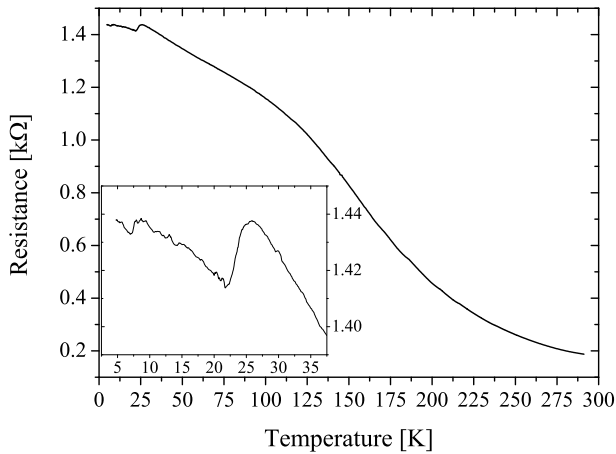}
\caption{\label{fig:R-T_C4} Resistance vs. temperature of a Ba-122\textbackslash TiO$_x$\textbackslash Pb junction with $7\,\mu$m~$\times~7\,\mu$m area and $d$(TiO$_x$)\,=\,2.0\,nm. The inset shows the same curve in the low temperature range.}
\end{figure}

By increasing the thickness of the TiO$_x$ barrier from 1.5\,nm the junction behavior changes dramatically. For a thickness of 2.0\,nm some of the junctions on the same substrate still show Josephson effects while others do not. For even higher thicknesses we could not observe Josephson effect at all. In fig.~\ref{fig:R-T_C4} the temperature dependence of the junction resistance is shown for a junction with a 2.0\,nm barrier without showing a Josephson effect. It can be seen that the resistance increases with decreasing temperature. These results infer that these junctions form either SIS' or an SINS' type ones. Also the superconducting transitions of the electrodes have only small (Ba-122) or quasi no (Pb) influence on the measured resistance, which is positive for the problem of spreading resistance.\cite{Doering2013b}
\begin{figure}[htbp]
\centering
\includegraphics[width=.95\columnwidth]{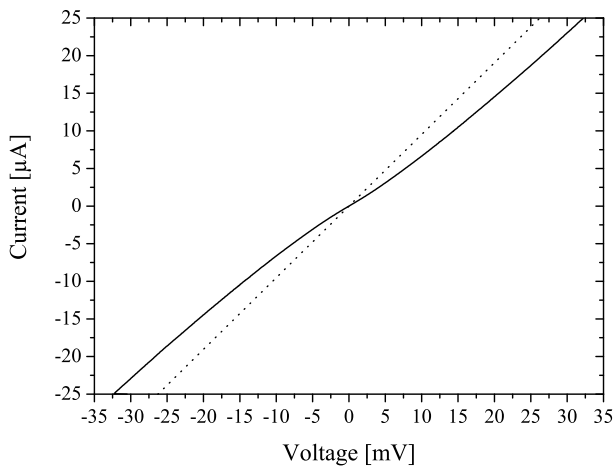}
\caption{\label{fig:I-V_C4} Current vs. Voltage characteristics of the same junction like in fig.~\ref{fig:R-T_C4} at $T$\,=\,4.2\,K (Solid line). The dotted line shows the ohmic line of $R$\,=\,1.05\,k$\Omega$.}
\end{figure}
\begin{figure}[htbp]
\centering
\includegraphics[width=.95\columnwidth]{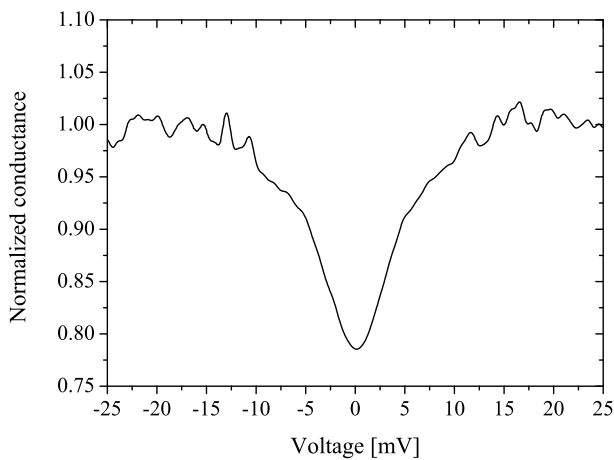}
\caption{\label{fig:G_C4} Normalized differential conductance vs. voltage of the junction used in fig.~\ref{fig:I-V_C4}.}
\end{figure}
In fig.~\ref{fig:I-V_C4} the corresponding $I$-$V$-characteristic at $T=4.2$\,K is shown and compared to the normal state resistance of 1.05\,k$\Omega$. In contrast to fig.~\ref{fig:V-I} it shows no Josephson effect but more tunnel-like behavior. This is supported by the conductance spectrum shown in fig.~\ref{fig:G_C4}. It can be seen, that there is no central peak but a clear dip down to 80\,percent of the normal state value. Additionally there are is only a small and noisy rise of the conductance beside the central dip. Such rise should be expected from energy dependence of the superconducting density of states (DoS). This behavior can be explained by two facts. Firstly, a high Dynes-parameter $\Gamma$, which includes scattering processes into the BTK-model \cite{Dynes1978,Plecenik1994} and lowers features compared to pure ballistic junctions. Secondly, spin-polarization of the pnictide should be taken into account. In such an extended BTK-model, the DoS side peaks can be lowered or vanish completely for present spin-polarization for any type of junction barrier. \cite{Mazin2001,Auth2003}\\
The critical thickness of the barrier for junctions with a pnictide base electrode is about 1\,nm lower than for junctions with the niobium base electrode. The possible reason for this is the lower roughness of the Ba-122 surface. While Ba-122 is an epitaxially grown thin film with nearly atomic flat surface, the niobium electrode is prepared from an e-beam evaporated layer on a Silicon wafer. Thus the niobium is amorphous with higher surface roughness than the Ba-122 thin film. 

\section{Summary}
To summarize, we used TiO$_x$ barriers firstly on junctions with conventional metal superconductors and showed that we were able to switch between Josephson effect and tunnel-like behavior by increasing the thickness of the barrier. Subsequently, we transferred the technology to junctions with iron pnictide base electrode and Pb counter electrode. For a double barrier with very thin Au and 1.5\,nm TiO$_x$ we could increase the resulted $I_cR_n$-product of SNS'-like Josephson junctions by at least a factor of 5 compared to former junctions with a pure 5\,nm gold barrier. Additionally, we are able to prepare high resistant SIS' or SINS'-like junctions for tunneling spectroscopy by further increasing the barrier thickness. With this first junctions Thus we were able to increase the properties and the possibilities of hybrid junctions with a pnictide electrode.

\acknowledgements

This work was partially supported by the DFG within SPP 1458 (project nos. SE664/15-2 and HA5934/3-1) and GRK 1621 and the European community under project IRON-SEA (project no. FP7-283141). Additionally, S.S. was funded by the Landesgraduiertenf\"{o}rderung Th\"{u}ringen. We would like to thank R. Nawrodt for his useful comments.

\section*{References}

\end{document}